----------------------------- Start of body part 4

\documentstyle[11pt]{article}

\begin{document}
\begin{titlepage}

\title{New Integrable Deformations of higher spin
Heisenberg-Ising chains}

\author{C\'esar G\'omez\thanks{Permanent address: Instituto de F\'{\i}sica
Fundamental, CSIC
Serrano 123. Madrid, SPAIN. e--mail addresses:
GOMEZC@SC2A.UNIGE.CH // IFFGOMEZ@ROCA.CSIC.ES} \\
{\it Dpartement de Physique Thorique} \\
{\it Universit de Gen
and \\ Germ\'an Sierra\thanks{e--mail address:SIERRA@CC.CSIC.ES }
\\ {\it Instituto de F\'{\i}sica Fundamental, CSIC} \\
{\it Serrano 123. Madrid, SPAIN}}

\date{}
\maketitle

\begin{abstract}
We show that the anisotropic Heisenberg-Ising chains with higher
spin allow, for special values of the anisotropy, integrable
deformations intimately related to the theory of quantum groups
at roots of unity. For the spin one case we construct and study
the symmetries of the hamiltonian which depends on a spectral variable
belonging to an elliptic curve. One of the points of this curve
yields the Fateev-Zamolodchikov hamiltonian of spin one and anisotropy
$\Delta = \frac{ q^2 + q^{-2}}{2} $ with $q$ a cubic root of unity.
In some other special points the spin degrees of freedom as well
as the hamiltonian splits into pieces
governed by a larger symmetry.
\end{abstract}


\vskip-16.0cm
\rightline{{\bf IFF-2/92}}
\rightline{{\bf March 1992}}
\vskip2cm

\end{titlepage}

One dimensional spin chains constitute an excellent laboratory
for the study of some non perturbative phenomena as for instance
 the dynamical mass generation in non linear $\sigma-$models [1].
Recently new experiments with NENP materials strongly
indicate the existence of a mass gap in $1D$ Heisenberg antiferromagnets
with integer spin [2], calling for a deeper understanding of
these systems.
Integrable one dimensional spin chains with spin $S$ higher than
$1/2$ were first formulated by Babujian, Takhtajan and others [3] for
the isotropic case and later on generalized to the anisotropic
case of spin one by Fateev and Zamolodchikov [4]. These
integrable spin chains are gapless and they correspond ,for
the periodic chains and a certain range of the anisotropy [5], to
Wess-Zumino-Witten models with central extension $\frac{3S}{S +
1}$ [6].

In this letter we define a new class of integrable spin chains
which can be viewed as integrable deformations of the
Heisenberg-Ising antiferromagnets with higher spin. These new spin
chains are built up by a procedure similar to the one used in the
construction of the Chiral Potts models [7], being in this sense
new descendents of the $6$ vertex models.
The mathematics which underlines these new chains is intimately
connected with the representation theory of quantum groups at
roots of unity [8]. We shall study the simplest case of the
quantum deformation of the affine extension of $Sl(2)$ namely
$U_q ( \widehat{Sl(2)} )$ with $q = \epsilon$ an $N-$th root of unity
( ${\epsilon}^N = 1 $).

The transfer matrix $T(u;q)$ of the $6$-vertex model is obtained
from the quantum $R$ matrix of $U_q (\widehat{Sl(2)})$ in the spin $1/2$
representation and it is given by [9]:

$$
R^{1/2, 1/2} (u,q) = sinh \left[ u + i\gamma \frac{1}{2} (1 + \sigma^Z
\otimes \sigma^Z) \right] + i \; sin\gamma  \; (\sigma^+ \otimes
\sigma^- + \sigma^- \otimes \sigma^+)
\eqno{(1)}
$$

\noindent
with $q = e^{ i \gamma}$.
The associated $1D$ hamiltonian is the well known
$XXZ$-hamiltonian with anisotropy parameter $\Delta = \frac{q + q^{-1}}{2}$,
which is the logarithmic derivative of the transfer matrix at
the point $u = 0$ :

$$
H_{XXZ} = i \; \frac{\partial}{\partial u} \; \ln T(u;q)
{\mid}_{u=0} = \sum_i \; {\sigma}^X_i {\sigma}^X_{i+1}
+ { \sigma}^Y_i {\sigma}^Y_{i+1}
+ \Delta \; { \sigma}^Z_i {\sigma}^Z_{i+1}
\eqno{(2)}
$$

A descendent of the $6-$vertex model is a model characterized by
a monodromy matrix $L(u)$ which satisfies the integrability equation:

$$
R^{1/2, 1/2} (u-v,q) \left( L_1 (u) \otimes L_2(v)
\right) = \left( L_2 (v) \otimes L_1(u) \right)
R^{1/2, 1/2} (u-v,q)
\eqno{(3)}
$$

\noindent
where $L_1(u)= L(u) \otimes \bf{1}   $ and $L_2(v)= \bf{1} $
$\otimes L(v)$.

Let us supose that we find two solutions $L^{\rho_1}$ and $L^{\rho_2}$
of (3) , the next step is to look for a new $R$-matrix
$R^{\rho_1 \rho_2}$ satisfying the Yang-Baxter equation and the condition:

$$
R^{\rho_1 \rho_2} (u-v) \left( L^{ \rho_1 }_1(u)
\otimes L^{ \rho_2 }_2 (v) \right) = \left( L^{ \rho_2 }_2(v)
\otimes L^{ \rho_1 }_1 (u) \right) R^{\rho_1 \rho_2} (u-v)
\eqno{(4)}
$$

The solutions to (3) are in one to one correspondence with the irreducible
representations of $U_q \widehat{ (Sl(2)})$ [10] and the $R-$matrix
solutions to (4) just provide the intertwiner for the
representations $\rho_1 $ and $\rho_2$. As an application of
this method
one could find the higher spin Heisenberg models of references [3-6]
by taking  $\rho_1 $ and $\rho_2$ to be integer or half integer
spin irreps of  $U_q  \widehat{(Sl(2)})$.
When $q = \epsilon$ is a
$N-$th root of unity a new class of finite dimensional irreps emerges.
The dimension of these irreps is equal to $N'$ where
$N'= N$ if $N$ is odd and $N'= N/2$ if $N$ is even [8]. In the
standard Chevalley basis the six generators of  $U_q \widehat{
(Sl(2)})$ can be given as:

$$
E_0 = x \;   F \;, \; F_0 = \frac{1}{x} E \;, \; K_0 = K^{-1}
\eqno{(5)}
$$

$$
E_1 = x \;  E \; , \; F_1 = \frac{1}{x} F \;, \; K_1 = K
$$

\noindent
with $E,F$ and $K$ satisfying the algebraic relations of
$U_q (Sl(2))$. The spectral parameter $x$ in (5) reflects the affine
nature of the representation.

The characteristic feature of quantum groups at
roots of unity is that the center of the algebra increases drastically
containing, in the case of $U_q (Sl(2))$ and in addition to the usual casimir,
the elements $E^{N'}, F^{N'}, K^{N'}$. The eigenvalues of these
central elements label
the new representations which can be classified into
cyclic, semicyclic and nilpotent. The cyclic irreps have non
vanishing values of $E^{N'}, F^{N'}, K^{N'}$ [11], the semicyclic
have $E^{N'}=0$ and $F^{N'}\neq 0$ (or viceversa) and the nilpotent
yield vanishing values of $E^{N'}$ and $F^{N'}$ while $ K^{N'}$
remains generic.

The solution to equations (3) and (4) associated to cyclic
irreps [12] yields the well known Chiral Potts models of references
[13] and they enjoy a $Z_N$ invariance symmetry absent in the higher
spin Heisenberg models. Some different models based on
semicyclic and nilpotent representations have been recently proposed
in references [14,15]. It is the purpose of this letter to
discuss new aspects of the model associated to nilpotent representations.
As we have said above a nilpotent irrep of  $U_q \widehat{ (Sl(2)})$
is labelled by the pair $(x,\lambda)$ where $K^{N'}={ \lambda}^{N'}$,
then the solution to (4) for two irreps $(x_1,\lambda_1)$and
$(x_2 ,\lambda_2)$ is given by [15]:

$$
R^{\lambda_1 \lambda_2} (u)^{l, r_1 + r_2 - l}_{r_1 r_2} =
\frac{1}
{\prod^{r_1+r_2-1}_{j=0} (e^u \lambda_1 \lambda_2
\epsilon^{-j} - e^{-u} \epsilon^j)} \times
$$

$$
\times {\sum^{r_1}_{l_1 = 0}
\sum^{r_2}_{l_2 = 0}} \left[ \begin{array}{c} r_1
\\ l_1 \end{array} \right] \left[ \begin{array}{c} r_2 \\ l_2
\end{array} \right] \frac{[l] ! [r_2 - l_2] !}{[r_1 + l_2] !
[r_2]!} (\epsilon - \epsilon^{-1})^{r_1 - l_1 +l_2}
$$

$$
\times \prod^{r_1 + l_2 -1}_{j= r_1} d_j (\lambda_1) \prod^{r_1
+l_2-1}_{j=l_1 +l_2} d_j (\lambda_1) \prod^{r_2
- 1}_{j=r_2 - l_2} d_j (\lambda_2) \prod^{r_1+r_2
-l-1}_{j=r_2 -l_2} d_j (\lambda_2)
$$

$$              \times \lambda^{l_2}_1 \lambda^{r_1-l_1}_2 \prod^{r_2 -l_2
-1}_{j=0} (e^u \lambda_2 \epsilon^{-j} - e^{-u} \lambda_1
\epsilon^j) \prod^{l_1 -1}_{j=0} (e^u \lambda_1 \epsilon^{-j+r_2-l_2}
-e^{-u} \lambda_2
\epsilon^{j+l_2 -r_2})
\eqno{(6)}
$$

\noindent
where $r_1,r_2,l$ and $r_1 + r_2 - l =0,1,..,N'$ and:

$$
e^u = x_1 / x_2
$$

$$
[n] = \frac{q^n - q^{-n}}{q - q^{-1}}
\eqno{(7)}
$$

$$
d^2_j (\lambda) = [j + 1]
\frac{\lambda {\epsilon}^{-j} - {\lambda}^{-1} \epsilon^j}
{\epsilon - \epsilon^{-1}}
$$

\noindent
with the following conventions: a) whenever in above products the upper index
is less than the lower index the result is one.
b) the constraint $l_1 + l_2 = l$ must be used to carry out
in the summation.

The $R-$matrix (6) can be used to define the Boltzmann weights
of an integrable vertex model with transfer matrix:

$$
T_{{\lambda}_0}(\lambda , u) = tr_a
\left( R^{\lambda {\lambda}_0}_{a L }(u)
R^{\lambda {\lambda}_0}_{a L-1 }(u) \cdots
R^{\lambda {\lambda}_0}_{a 1 }(u) \right)
\eqno{(8)}
$$

\noindent
where the trace is taken over the auxiliary space denoted by $a$
which carries the representation $\lambda$,
while the irrep ${\lambda}_0$ is
associated to the quantum space where the transfer matrix is acting.
The Yang-Baxter relation for $R^{\lambda_1 \lambda_2} (u)$
implies an integrability relation for the transfer matrix involving
both variables $u$ and $\lambda$:

$$
[T_{{\lambda}_0}(\lambda , u),T_{{\lambda}_0}(\lambda' , u')] = 0
\eqno{(9)}
$$

The spin chain hamiltonian that follows from the transfer matrix
(8) is defined as usual:

$$
H({\lambda}_0) = i \; \frac{\partial}{\partial u} \ln
T_{{\lambda}_0}({\lambda}_0 , u) \mid_{u =0}
\eqno{(10)}
$$

The two parameter integrability equation (9) implies the existence
of a new local conserved quantity $Q ({\lambda}_0)$ :

$$
Q({\lambda}_0) = 2 \; \lambda \frac{ \partial}{ \partial \lambda}
\ln T_{{\lambda}_0}(\lambda , u =0) \mid_{\lambda = {\lambda}_0}
\eqno{(11)}
$$

\noindent
which satisfy :

$$
[ Q({\lambda}_0), H({\lambda}_0)] = 0
\eqno{(12)}
$$

The hamiltonian (10) that follows from the $R-$ matrix (6) in
the case ${\epsilon}^3 = 1$ is after a lengthly calculation
given by:

$$
\hat{H}({\lambda}_0) = \sum^{L}_{k=1}
\frac{{\lambda}_0 \epsilon + {\lambda}^{-1}_0 {\epsilon}^{-1}}{2}
( S^X_k S^X_{k+1} + S^Y_k S^Y_{k + 1} ) - \frac{1}{2}
S^Z_k S^Z_{k + 1}
$$

\[
- ( S^X_k S^X_{k+1} + S^Y_k S^Y_{k + 1} )^2 + \frac{1}{2}
(S^Z_k S^Z_{k + 1})^2
\]

$$
+ \left( \frac {{\lambda}_0 \epsilon + {\lambda}^{-1}_0 \epsilon^{-1}}{2}
+ \omega \right) \;
\left[  ( S^X_k S^X_{k+1} + S^Y_k S^Y_{k + 1} )S^Z_k S^Z_{k + 1}
+ \leftrightarrow \right]
\eqno{(13)}
$$

$$
-  \frac{3}{2}
\left( (S^Z_k)^2 + ( S^Z_{k + 1})^2 \right)
- \frac {{\lambda}_0 \epsilon - {\lambda}^{-1}_0 {\epsilon}^{-1}}
{2 (\epsilon - {\epsilon}^{-1}) }
( S^X_k S^X_{k+1} + S^Y_k S^Y_{k + 1} )(S^Z_k +S^Z_{k + 1})
$$

$$
+ \frac {{\lambda}^2_0 {\epsilon}^{-1} - {\lambda}^{-2}_0
{\epsilon}}
{2 (\epsilon - {\epsilon}^{-1}) }
\left( S^Z_k +S^Z_{k + 1} \right)
$$

\noindent
where in order to avoid singularities we have redefined the hamiltonian
(10) as $\hat{H}({\lambda}_0) =   \frac{i}{2}
( \epsilon - \epsilon^{-1}) \omega^2 (\lambda_0) H(\lambda_0)$
with  $ {\omega}({\lambda})=
\sqrt{
( \lambda - {\lambda}^{-1} )( \lambda {\epsilon}^{-1} -
{\lambda}^{-1} \epsilon )}/ i (\epsilon
- \epsilon^{-1}) $. The matrices appearing in (13)
are the spin 1 matrices of $SU(2)$.
We may ask at this stage what is the relation between this hamiltonian
and the Fateev-Zamolodchikov hamiltonian which describes the anisotropic
Heisenberg model of spin 1, whose explicit expresion reads [4]:

$$
H_{FZ}(q) = \sum^{L}_{k = 1}
S^X_k S^X_{k+1} + S^Y_k S^Y_{k + 1}
+ \frac{ q^2 + {q}^{-2}}{2}
S^Z_k  S^Z_{k + 1}
$$

$$
-( S^X_k S^X_{k+1} + S^Y_k S^Y_{k + 1} )^2
- \frac{ q^2 + {q}^{-2}}{2}
(S^Z_k  S^Z_{k + 1})^2
\eqno{(14)}
$$

$$
+ (1 - q - q^{-1})
\left[  ( S^X_k S^X_{k+1} + S^Y_k S^Y_{k + 1} )S^Z_k S^Z_{k + 1}
+ \leftrightarrow \right]
$$

$$
+ \frac{ q^2 + {q}^{-2}- 2}{2}
[(S^Z_k)^2 +(  S^Z_{k + 1})^2 ]
$$

Compairing (13) and (14) we arrive at the result (choosing the branch
$\omega( \lambda_0 = \epsilon^2) = 1 )$:

$$
\hat{H}({\lambda}_0) \mid_{{\lambda}_0 = \epsilon^2} =
H_{FZ}(q) \mid_{q = \epsilon, {\epsilon}^3 = 1}
\eqno{(15)}
$$

This result is also consistent with the fact that $\lambda_0 =\epsilon^2$
corresponds to a representation of spin 1.
If we move away from the point ${\lambda}_0 = {\epsilon}^2$ in the
$\lambda$ direction we obtain an integrable deformation of the
FZ hamiltonian. It is known that the FZ model is gapless and it corresponds
to a WZW model with central extension 3/2 [5, 6]. The
thermodynamic properties of the model based on (13)
are under investigation but we believe that the deformation in the
$\lambda$ direction corresponds to a relevant perturbation.
One more thing we can add from our study is that the FZ
hamiltonian at the value $q = \epsilon$ has a new local conserved
operator $Q(\lambda)$, as can be seen from eq.(12). Thus we are
tempted to conjecture that the existence of new local conserved quantities
is the signal for the existence of integrable deformations of
the kind discussed in this letter.

In order to study the symmetries of the hamiltonian (13) we
shall use the matrices $ X= \frac{1}{\sqrt{2}}(S_X + i S_Y) $
and $Z=S_Z$, then (13) and the conserved quantity  $\hat{Q}({\lambda}_0)
=   \frac{i}{2}
( \epsilon - \epsilon^{-1}) \omega^2 (\lambda_0) Q(\lambda_0)$
read:

$$
\hat{H}(\lambda)= \sum^{L}_{k=1} \; \alpha
(X_k Z_k Z_{k+1} X^{\dagger}_{k+1} + Z_k X^{\dagger}_k X_{k+1} Z_{k+1})
$$

$$
+ \beta
(Z_k X_k X^{\dagger}_{k+1} Z_{k+1} + X^{\dagger}_k Z_k Z_{k+1} X_{k+1})
+ \rho ( X^2_k (X^{\dagger}_{k+1})^2 + (X^{\dagger}_k)^2 X^2_{k+1})
\eqno{(16.a)}
$$

$$
+ \omega
(Z_k X_k Z_{k+1} X^{\dagger}_{k+1} + Z_k X^{\dagger}_k Z_{k+1} X_{k+1}
+ X_k Z_k X^{\dagger}_{k+1} Z_{k+1} + X^{\dagger}_k Z_k X_{k+1} Z_{k+1})
$$

$$
+ \frac{ \alpha^2 - \beta^2}{\rho} ( 1 - Z_k) +
\rho ( Z^2_k -1 )
$$

$$
\hat{Q}(\lambda)=  -i \sum^{L}_{k=1} \; \alpha
(X_k Z_k Z_{k+1} X^{\dagger}_{k+1} - Z_k X^{\dagger}_k X_{k+1} Z_{k+1})
$$

$$
+ \beta
(Z_k X_k X^{\dagger}_{k+1} Z_{k+1} - X^{\dagger}_k Z_k Z_{k+1} X_{k+1})
+ \rho ( X^2_k (X^{\dagger}_{k+1})^2 - (X^{\dagger}_k)^2 X^2_{k+1})
\eqno{(16.b)}
$$

$$
+ \omega
(Z_k X_k Z_{k+1} X^{\dagger}_{k+1} - Z_k X^{\dagger}_k Z_{k+1} X_{k+1}
+ X_k Z_k X^{\dagger}_{k+1} Z_{k+1} - X^{\dagger}_k Z_k X_{k+1} Z_{k+1})
$$

The parameters $(\alpha,\beta,\rho,\omega)$ that we shall collectively
denote by $\xi$ are given by:

$$
\alpha = - (\lambda - \lambda^{-1})/(\epsilon - \epsilon^{-1})
$$

$$
\beta   = (\lambda \epsilon^{-1} - \lambda^{-1} \epsilon)/
(\epsilon - \epsilon^{-1})
\eqno{(17)}
$$

$$
\rho = -1
$$

$$
\omega = \sqrt{(\lambda - \lambda^{-1})(\lambda \epsilon^{-1} -
\lambda^{-1} \epsilon)}/ i (\epsilon - \epsilon^{-1})
$$

\noindent
more generally, considering $\xi$ as a coordinate in $CP^3$
we obtain an algebraic curve of genus 1 defined by the equations:

$$
\alpha \beta  = \omega^2
$$

$$
\alpha^2 + \beta^2 - \alpha \beta  = \rho^2
\eqno{(18)}
$$

\noindent
hence the parameter $\lambda$ in (17) plays the role of a uniformization
variable.

The hermiticity regions of the hamiltonian (13) or (16.a) for
$\epsilon = e^{ 2 \pi i /3}$ corresponds to the values of
$\lambda = e^{i \phi}$ lying in the intervals
$ \phi \in [0,  2\pi/3] \cup [\pi, 5\pi/3]$, which coincide with
two homology cycles, say a-cycles, of the elliptic curve (18).The
non-contractility of these hermiticity regions is a potential
source for non trivial Berry phases.

Among the authomorphisms of the curve (18) we shall distingish three
of them which relate hamiltonians with the same spectrum. They are
given by:

$$
\begin{array}{ccc}
type \; I & type \; II & type \; III \\
\alpha \rightarrow -\alpha & \alpha \rightarrow \alpha & \alpha
\rightarrow \beta  \\
\beta  \rightarrow - \beta  & \beta  \rightarrow \beta  & \beta
\rightarrow \alpha \\
\omega \rightarrow -\omega & \omega \rightarrow - \omega &
\omega \rightarrow \omega \\
\rho \rightarrow \rho & \rho \rightarrow \rho & \rho \rightarrow
\rho
\end{array}
\eqno{(19)}
$$

The corresponding unitary transformation is given by:

$$
H(g(\xi))= U_g H(\xi) U^{\dagger}_g
\eqno{(20)}
$$

\noindent
where

$$
\begin{array}{lcr}
U_I = \prod^L_{k=1} e^{i k \pi Z_k}, & U_{II} = \prod^L_{k = 1}
e^{ i \frac{\pi}{2} Z^2_k }, & U_{III} = \prod^L_{k=1} C_k
\end{array}
\eqno{(21)}
$$
and $C$ is the charge conjugation matrix:

$$
C =
\left( \begin{array}{ccc}
0 & 0 & 1 \\
0 & 1 & 0 \\
1 & 0 & 0
\end{array}
\right)
\eqno{(22)}
$$
\noindent
Authomorphism of type III is equivalent to the charge conjugation
operation :$\lambda \rightarrow     \epsilon
\lambda^{-1}$. As we see from (17) the FZ model is selfconjugated,
while the points $\lambda = \pm 1 $ and $\lambda = \pm \epsilon$
are interchanged under conjugation.These points, which correspond
to the branch points of the elliptic curve (18), are very special
from the point of view of the representation theory of $U_{\epsilon}(Sl(2))$
[16].In fact when  $\lambda = \pm 1 $ or $\lambda = \pm \epsilon$
the three dimensional representation of  $U_{\epsilon}(Sl(2))$ with
$\epsilon^3 =1$ breaks into a singlet and a doublet and the corresponding
hamiltonian (13) takes a particular simple form:

$$
H(\lambda= 1)=    \sum^L_{k= 1}
-(\tau^+_k \tau^-_{k+1} + \tau^-_k \tau^+_{k+1} +
\rho^+_k \rho^-_{k+1} + \rho^-_k \rho^+_{k+1}) + \sigma^0_k
\eqno{(23.a)}
$$

$$
H(\lambda= \epsilon )=    \sum^L_{k= 1}
-(\sigma^+_k \sigma^-_{k+1} + \sigma^-_k \sigma^+_{k+1} +
\rho^+_k \rho^-_{k+1} + \rho^-_k \rho^+_{k+1}) + \tau^0_k
\eqno{(23.b)}
$$

\noindent
where the matrices $\sigma, \tau$ and $\rho$ yield three
possible embeedings of $U(2)$ inside $U(3)$ :

$$
\sigma^+ = \left( \begin{array}{ccc}
0 & 0 & 0 \\
0 & 0 & 1 \\
0 & 0 & 0
\end{array}
\right) \;
\sigma^- = \left( \begin{array}{ccc}
0 & 0 & 0 \\
0 & 0 & 0 \\
0 & 1 & 0
\end{array}
\right) \;
\sigma^Z = \left( \begin{array}{ccc}
0 & 0 & 0 \\
0 & 1 & 0 \\
0 & 0 & -1
\end{array}
\right) \;
\sigma^0 = \left( \begin{array}{ccc}
0 & 0 & 0 \\
0 & 1 & 0 \\
0 & 0 & 1
\end{array}
\right) \;
$$

$$
\tau^+ = \left( \begin{array}{ccc}
0 & 1 & 0 \\
0 & 0 & 0 \\
0 & 0 & 0
\end{array}
\right) \;
\tau^- = \left( \begin{array}{ccc}
0 & 0 & 0 \\
1 & 0 & 0 \\
0 & 0 & 0
\end{array}
\right) \;
\tau^Z = \left( \begin{array}{ccc}
1 & 0 & 0 \\
0 &-1 & 0 \\
0 & 0 & 0
\end{array}
\right) \;
\tau^0 = \left( \begin{array}{ccc}
1 & 0 & 0 \\
0 & 1 & 0 \\
0 & 0 & 0
\end{array}
\right) \;
\eqno{(24)}
$$

$$
\rho^+ = \left( \begin{array}{ccc}
0 & 0 & 1 \\
0 & 0 & 0 \\
0 & 0 & 0
\end{array}
\right) \;
\rho^- = \left( \begin{array}{ccc}
0 & 0 & 0 \\
0 & 0 & 0 \\
1 & 0 & 0
\end{array}
\right) \;
\rho^Z = \left( \begin{array}{ccc}
1 & 0 & 0 \\
0 & 0 & 0 \\
0 & 0 & -1
\end{array}
\right) \;
\rho^0 = \left( \begin{array}{ccc}
1 & 0 & 0 \\
0 & 0 & 0 \\
0 & 0 & 1
\end{array}
\right)   \;
$$

\noindent
Thus for $\lambda = 1$ the spin variables of the chain
$(e_0, e_1, e_2)$ break into a singlet $ e_0$ and a doublet
$(e_1, e_2)$. The hamiltonian (23.a) describes a XX-interaction between
the singlet and the two components of the doublet and it has
a $SU(2)\otimes U(1)$ symmetry generated by the
operators $\sum_k \sigma^{\mu}_k $ with $\mu = 0, z, +,- $ .The same
situation occurs for the case $\lambda = \epsilon$ where now the
doublet is formed by the states $e_0 $ and $e_1$ while the
singlet is given by $e_2$.
Outside the points $\lambda = \pm1, \pm \epsilon$ the symmetry
$SU(2) \otimes U(1)$ is brooken down to a $U(1)$ symmetry, that of
the Cartan part of $U_{\epsilon}(Sl(2))$.

A possible physical interpretation of the hamiltonians (23)
is in terms of the Hubbard model with infinite Coulomb repulsion.
Indeed such a repulsive force forbids two electrons of opposite
spin to occupy the same lattice position, so that the effective
Hilbert space is isomorphic to the one of models (23). Under this
interpretation, the $SU(2) \otimes U(1)$ symmetry is nothing else
but the rotation group symmetry times the total electron charge.

It may be interesting to observe that the hamiltonian (23.a) can
be also obtained through the following "flavor" generalization
of the 6-vertex R-matrix:

$$
R^{0 0}_{0 0} = sinh(u + i \gamma)
$$

$$
R^{i 0}_{j 0} = R^{0 i}_{0 j} = \delta^i_j \; sinh( u)
\eqno{(25)}
$$

$$
R^{i 0}_{0 j} = R^{0 i}_{j 0} = i \delta^i_j \; sin \gamma
$$

$$
R^{i k}_{j l} = \delta^i_j \; \delta^k_l
\; sinh(u + i \gamma)
$$

\noindent
where $i= 1,2$ and $\gamma = \frac{\pi}{2}$ ( the generalization
to $i= 1,2,..,\cal{N}$ with $\cal{N}$ and $\gamma$ arbitrary is
straightforward ).
The R-matrix (25) shows explicitely that the underlying symmetry
of the model at the points $\lambda = \pm1, \pm \epsilon$ is really
$U_{e^{i \pi/2}}(\widehat{Sl(2)}) \otimes SU(2) \otimes U(1)$.

The situation described above is not a peculiarity of the case of
$q$ being a cubic root of unit. In fact for a generic $N-$th root
of unit there are $2 (N'-1)$ special points: $\lambda = \pm1, \pm
\epsilon, \cdots,  \pm \epsilon^{N'-2}$,
where the $N'-$dimensional representation
breaks into two irreps. More concretely at the point $\lambda = \pm
\epsilon^{2 J}$ it breaks into a representation of spin $J$ and another
one with spin $ \frac{1}{2} N'- J - 1$.

To finish this letter we shall briefly comment on the Bethe
equations for the hamiltonian (13). These were derived using
standard Bethe ansantz techniques in reference [15]
and the result turns out to be :

$$
\left( \frac{ e^{u_j} \epsilon^{1/2} \lambda^{1/2}_0 -
e^{-u_j} \epsilon^{-1/2} \lambda^{-1/2}_0 }
{  e^{u_j} \epsilon^{1/2} \lambda^{-1/2}_0 -
e^{-u_j} \epsilon^{-1/2} \lambda^{1/2}_0 }
\right)^L = \prod^M_{k \neq j}
\frac{ sinh( u_j - u_k + i \gamma )}{ sinh( u_j - u_k - i \gamma )}
\eqno{(26)}
$$

\noindent
when $\lambda_0$ is replaced by $\epsilon^{2S}$, with $ S$
integer or half integer, equations (26) become the Bethe equations
of the integrable spin $S$-chain of reference [5]. At the
special points where the representation breaks into pieces a
"non abelian" or nested
generalization of the algebraic Bethe ansatz is needed,
which should take care of the extra symmetry that emerges. In the
simplest case we have discussed in this letter this generalization
can be extracted from the R-matrix (25).

We summarize our results in the following items: i) construction
of the hamiltonian of a new class of integrable 1D Heisenberg-Ising
chains, ii) interpretation of these models as integrable deformations
of WZW models at the microscopic level, iii) identification of the
parameter space of the hamiltonian of the cubic case
(i.e. $\epsilon^3  =1$) with an algebraic curve of genus 1 and iv)
discovery of spin chains with an enlarged non abelian symmetry
at the orbifold points of De Concini and Kac which seem to be closely
related to the Hubbard model with
an infinite Coulomb repulsive force.

\subsection*{Acknowledgements}

We would like to thank Alexander Berkovich for discussions and
for sharing with us his knowledge and insights.


\begin{thebibliography}{99}
\bibitem{16}  The points $\lambda = \pm 1,\;  \pm \epsilon $ for
              $\epsilon^3 = 1$ are special in the representation theory
              of $U_{\epsilon}(Sl(2))$, in the sense that they
              are fixed points under the action of the
              group of authomorphisms
              induced by the central Hopf subalgebra,
              which act on $Spec Z_{\epsilon}$ ( see
              ref.[8] for more details).
\end{thebibliography}
\end{document}